\documentclass[iop]{emulateapj}
\usepackage{amsmath}
\usepackage{amssymb}
\usepackage{epstopdf}
\usepackage{graphicx}
\usepackage{float}
\usepackage{natbib}
\usepackage{footnote}
\usepackage{algorithm,algorithmic}
\usepackage{bbold}

\slugcomment{submitted to APJ}

\shorttitle{Fast discrete Dispersion Measure Transform}
\shortauthors{Zackay \& Ofek}

\begin{document}


\title{An accurate and efficient algorithm for detection of radio bursts with an unknown dispersion measure, for single dish telescopes and interferometers}


\author{Barak Zackay\altaffilmark{} and Eran O. Ofek\altaffilmark{}}
\affil{Benoziyo Center for Astrophysics, Weizmann Institute of Science, 76100 Rehovot, Israel}

\email{bzackay@gmail.com}
\email{Eran.ofek@weizmann.ac.il}




\begin{abstract}
Astronomical radio bursts disperse while traveling through the interstellar medium.
To optimally detect a short-duration signal within a frequency band, we have to precisely compensate for the pulse dispersion, which is a computationally demanding task.
We present the Fast Dispersion Measure Transform (FDMT) algorithm for optimal detection of such signals.
Our algorithm has a low theoretical complexity of
$2N_fN_t+ N_tN_d\log_2(N_f)$ where $N_f$, $N_t$ and $N_d$ are the numbers of frequency bins, time bins,
and dispersion measure bins, respectively.
Unlike previously suggested fast algorithms our algorithm conserves the sensitivity of brute force dedispersion.
Our tests indicate that this algorithm, running on a standard desktop computer, and implemented in a high-level
programming language, is already faster than the state of the art dedispersion codes running on graphical processing units (GPUs). 
We also present a variant of the algorithm that can be efficiently implemented on GPUs.
The latter algorithm's computation and data transport requirements are similar to those of two-dimensional FFT,
indicating that incoherent dedispersion can now be considered a non-issue while planning future surveys.
We further present a fast algorithm for sensitive dedispersion of pulses shorter than normally allowed by incoherent dedispersion. In typical cases this algorithm is orders of magnitude faster than coherent dedispersion by convolution.
We analyze the computational complexity of pulsed signal searches by radio interferometers.
We conclude that, using our suggested algorithms, maximally sensitive blind searches for 
such pulses is feasible using existing facilities.
We provide an implementation of these algorithms in Python and MATLAB.

\end{abstract}



\section{Introduction}\label{sec:Introduction}
When a radio pulse propagates through the interstellar and intergalactic plasma, different frequencies travel at different speeds.
This phenomenon, known as dispersion, hinders the detection
of radio pulses.
This is because integrating over many frequencies during a given time frame dilutes the signal with noise, as only a single frequency contributes signal at any given interval within the integration frame. 
The solution to this problem is to dedisperse the signal
(i.e., to apply frequency dependent time delays to the signal prior to integration).
Since in most cases the dispersion is a priori unknown,
we need to test a large number of dispersions.
The best dispesrion is the one that maximizes the signal-to-noise
ratio of the pulse. A different way to look at this problem
is that we need to integrate the flux along many
dispersion curves in the frequency-time domain.

The difference in pulse arrival time between two frequencies is given by:
\begin{equation}
\Delta{t}=t_{2}-t_{1}= 4.15 {\rm DM} (f_{1}^{-2} - f_{2}^{-2})\,{\rm ms},
\label{eq:Dt}
\end{equation}
where ${\rm DM}$ is called the dispersion measure of the signal and it is traditionally measured in units of ${\rm pc}\,{\rm cm}^{-3}$. $f_{i}$ are frequencies measured in GHz and $t_{i}$ is the arrival time of the signal at frequency $f_{i}$.
For brevity, throughout the paper, we will use $d$ to denote
the dispersion measure in which all the dimensional constants
are absorbed, i.e., the relation is given by
\begin{equation}
\Delta{t}=t_{2}-t_{1}= d (f_{1}^{-2} - f_{2}^{-2}).
\label{eq:DtD}
\end{equation}

The raw input from a radio reciever is a time series of voltage measurments
sampled typically at high frequency (e.g., $\sim 100$\,MHz).
We denote the sampling interval by $\tau$.
In order to generate a spectrum ($I[t,f]$) as a function of time ($t$) and frequency ($f$) the time series is
divided to blocks of size $N_{f}$ samples, and each block is then Fourier transformed
(a process known as Short Time Fourier Transform, or STFT)
and the absolute value squared at each frquency is saved\footnote{Sometimes, an additional stage of binning is then applied to reduce the time resolution.}.

\begin{figure}[h]
\includegraphics[width = 86mm]{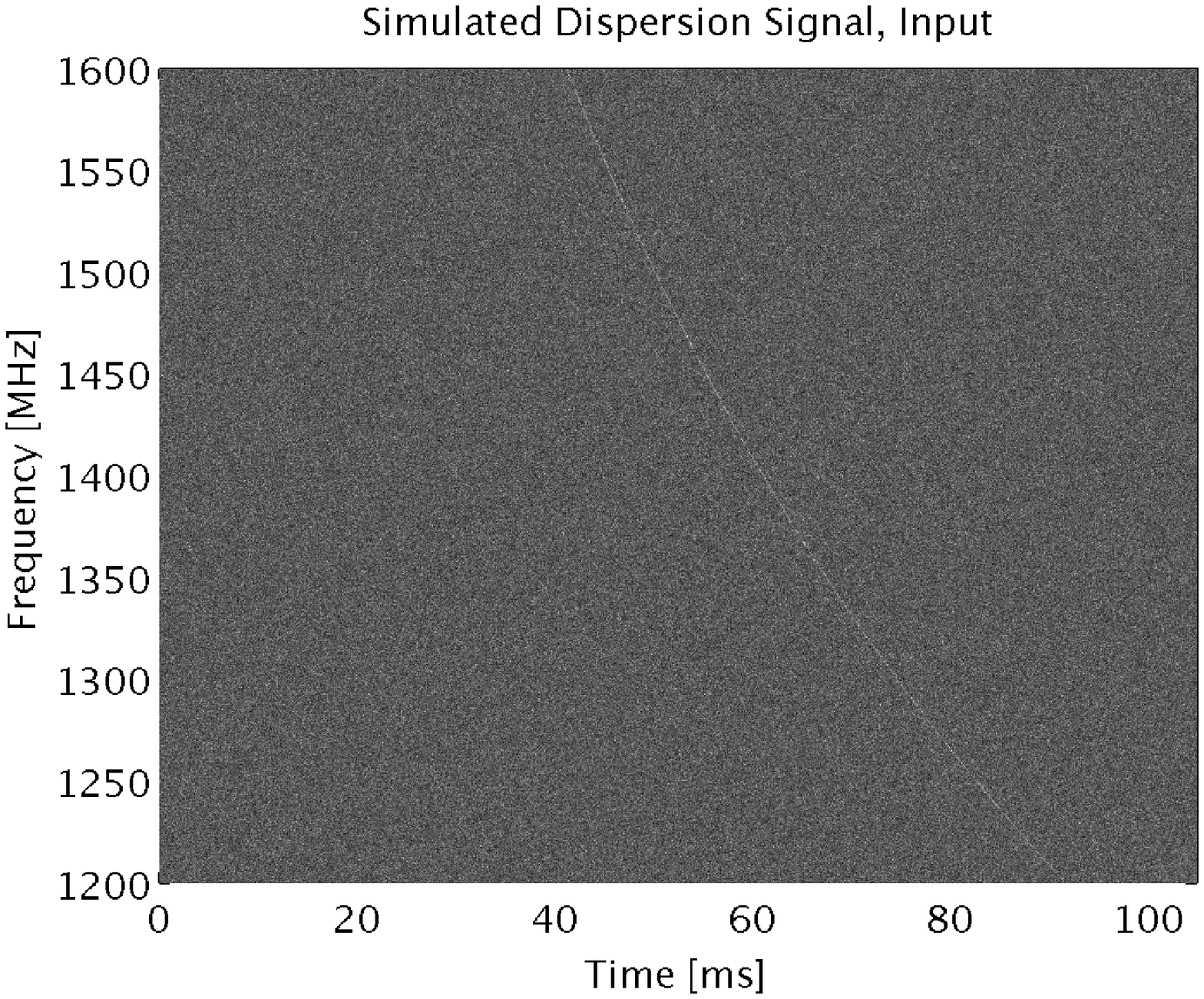}
\includegraphics[width = 86mm]{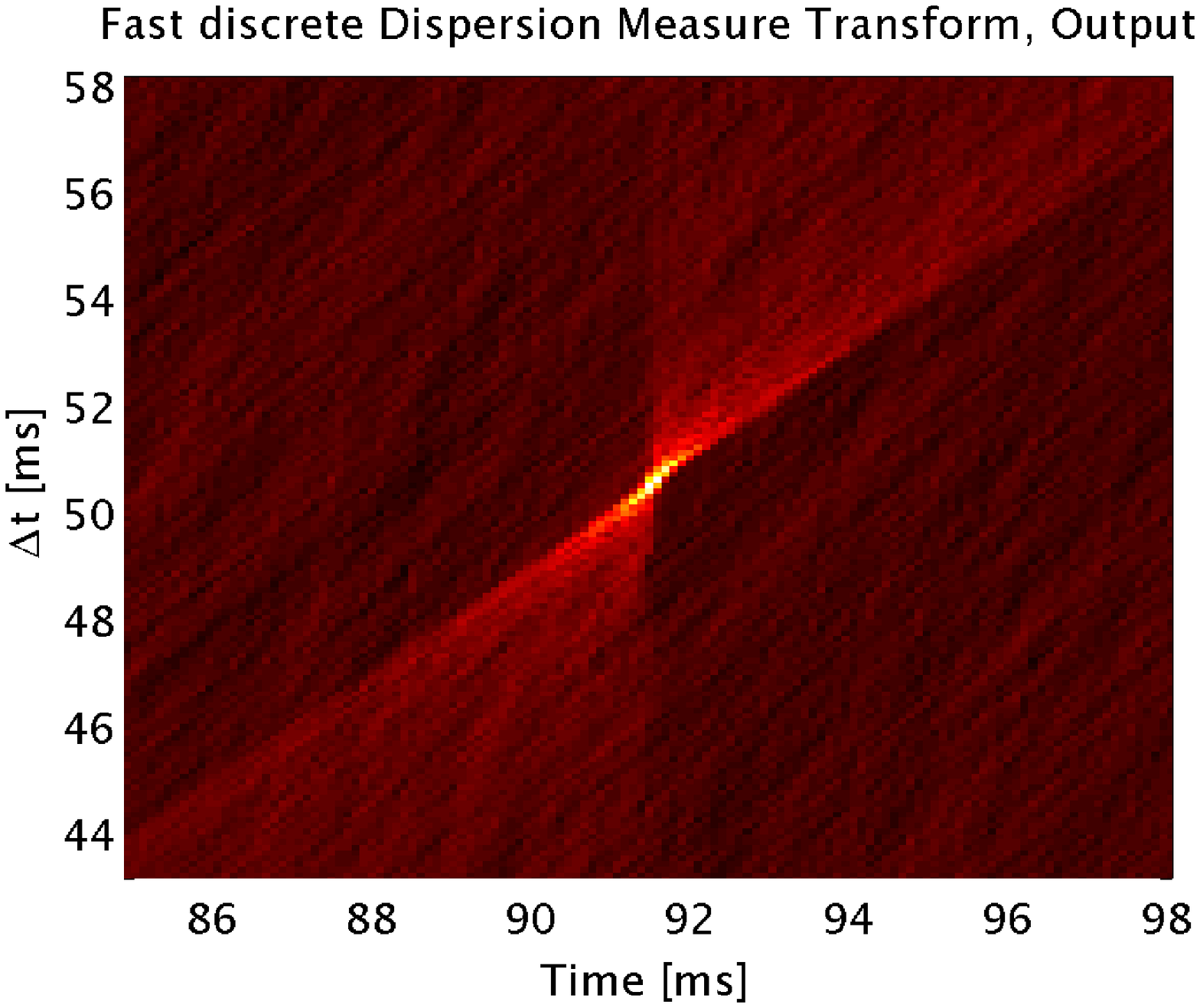}
\caption{A dispersed signal and its dispersion transform (FDMT), based on simulated data. The top panel shows a 0.1ms wide dispersed pulse with $D=40\, {\rm pc}\, {\rm cm}^{-3}$. The bottom panel shows a zoom in on the significant part of the dedispersion transform. Notice that the Y axis has units of time because the dispersion measure is parametrized by the total time delay between the pulses entrance and exit. Note that because the pulse is so thin, any slight error on the dispersion path will immediately result in significant loss of the pulse power.}\label{fig:InputOutput}
\end{figure}

There are two distinct processes that we can apply to dedisperse a signal: incoherent dedispersion and coherent dedispersion.
The term incoherent dedeispersion refers to applying frequency-dependent time delays to the $I(t,f)$ matrix,
while coherent dedispersion involves applying a frequency-dependent phase shift directly to the Fourier transform of the raw signal.
This subtle difference is important -- incoherent dedispersion is only an approximation that is valid
under certain conditions that we are going to review shortly (\S\ref{sec:SensitivityIncoherentDedispersion}).
A typical input matrix ($I[t,f]$) to incoherent dedispersion is presented in the top panel of Figure \ref{fig:InputOutput}, while on the bottom we show a zoom in on the output of the transform.
 
The exact mathematical description of signal dispersion is the multiplication of the Fourier transform of the raw signal with the phase-only filter \begin{equation}\hat{H}(f) = \exp{\left(\frac{2\pi id}{f+f_0}\right)}\end{equation}  \citep{HandbookOfPulsarAstronomy}
, where $f_0$ is the base-band frequency.
In order to exactly dedisperse the signal, we can apply the inverse of this shift. This process is referred as coherent dedispersion. 

The computational requirements of incoherent dedispersion are more tractable than those of coherent dedispersion, and therefore whenever a blind search for astrophysical pulses is done, incoherent dedispersion is usually the method of choice (e.g. \citealp{LOFARPulsar}). 
The main practical motivation for improving dedispersion algorithms is to allow efficient analysis of modern radio interferometers data.
When a blind search for new sources is conducted using a multi-component radio interferometer, coherent dedispersion of a large number of synthesized beams is the most sensitive detection method, but usually this is unfeasible. The standard solution for this problem is to either combine the power from all antennas incoherently, or to use only a small core of the interferometer for blind searches. Both approaches result in a significant loss of sensitivity and angular resolution that can amount to an order of magnitude sensitivity loss.
It is important to improve upon these methods, especially when searching for non-repeating radio transients such as fast radio bursts (FRBs; \citealp{LorimerBurst,FRBS}). Efficient detection of such objects requires both high sensitivity and a good spatial localization that is calculated in real time. This is crucial for the multi-wavelength followup of these illusive transients (e.g., \citealp{PetroffFRBsNullDetection}).

In this paper, we present the Fast discrete Dispersion Measure Transform algorithm (henceforth FDMT) to solve the problem of incoherent dedispersion. FDMT is a transform algorithm, having (generally) equal sizes for both input and output, and complexity that is only logarithmically larger than the input itself.

In addition, we present a hybrid algorithm that achieves both the sensitivity of coherent dedispersion, and the computational efficiency of incoherent dedispersion. Finally, we show that using this algorithm, it is feasible to perform blind searches with modern radio interferometers, and consequently to open new frontiers in the search for pulsars and radio transients.

The structure of the paper is as follows.
In \S \ref{sec:SensitivityIncoherentDedispersion} we analyze the sensitivity of incoherent dedispersion.
In \S \ref{sec:ExistingMethods} we review the existing approaches for incoherent dedispersion. 
In \S \ref{sec:FDMTAlgorithm} we describe the proposed algorithm, along with its complexity analysis.
In \S \ref{sec:FDMTFFT} we present a variant of the algorithm that utilizes the fast Fourier transform to make the algorithm much more parallel friendly.
In \S \ref{sec:RuntimeAndBenchmarking} we compare the runtime of the implementation we provide with existing implementations of brute force dedispersion.
In \S \ref{sec:CoherentIncoherentHybrid} we propose a new hybrid algorithm for detection of pulses shorter than sensitively detectable by incoherent dedispersion.
In \S \ref{sec:DedispersionInterferometers} we discuss the application of the proposed algorithms for interferometers, and show that sensitive detection of short pulses, with maximal resolution, using all the elements of the interferometer, is feasible with current facilities.
We conclude in \S \ref{sec:Conclusion}.


\section{Sensitivity analysis of incoherent dedispersion} \label{sec:SensitivityIncoherentDedispersion}

In this subsection, we develop the condition on the pulse length, the sampling interval and the dispersion delay that allows sensitive detection with incoherent dedispersion. This will be of importance in section \S\ref{sec:CoherentIncoherentHybrid}.  

We denote by $x$ the raw voltage signal, and by $N_s$ the total number of samples. 
We further denote the pulse duty time (length) by $t_p  =  N_p\tau$, where $N_p$ is the number of samples within the pulse.
The optimal score for pulse detection is the sum of the squared voltage measurements within the pulse start time ($t_0$) and end time, 
\begin{equation}
S_{\rm opt} = \sum_{j=0}^{N_p}{x(t_0+j)^2}.
\end{equation}
We further define the dispersion time $t_d$ to be the total delay of the pulse within the band and we define $N_d$ to be the dispersion time in units of samples, i.e. $t_d =N_d\tau$.

An important property of the dispersion kernel $H(f)$ is that it is power preserving ($|\hat{H}(f)|^2 = 1$).
Another important property is that the majority of pulse power will lay within the dispersion curve in $I(t,f)$.
By summing over the dispersion curve in  $I$, we also sum the power of the noise outside the pulse, which its total variance is proportional to the number of added $I$ bins--i.e., to the length of the dispersion curve.
The total length of the dispersion curve can be approximated by \begin{equation}\sim\max\left(1,\frac{{N_{p}}}{{N_{f}}}\right)\sqrt{{N_{f}}^2+\frac{{N_{d}}^2}{{N_{f}}^2}},\end{equation} assuming the dispersion curve is close to linear.

Therefore, the ratio of the noise power summed by incoherent dedispersion and the noise power within the pulse (assuming ${N_{p}}<{N_{f}}$) is \begin{equation}\frac{\sqrt{{N_{f}}^2+\frac{{N_{d}}^2}{{N_{f}}^2}}}{{N_{p}}}.\end{equation} 
Immediately, we get that the choice of ${N_{f}}$ that maximizes sensitivity is \begin{equation}
{N_{f}} = \sqrt{{N_{d}}},
\end{equation} regradless of pulse length ${N_{p}}$. In order for the sensitivity loss to be less than a factor of $\sqrt{2}$, we need both \begin{equation}{N_{d}}^2 < {N_{f}}^2{N_{p}}^2\end{equation} and \begin{equation}{N_{f}}^2 < {N_{p}}^2.\end{equation} 
This implies that \begin{equation}{N_{d}} < {N_{p}}{N_{f}} < {N_{p}}^2.\end{equation}
This transforms to an important relation between the minimal pulse duration $t_p$, the maximal dispersion time $t_d$, and the sampling time $\tau$.
\begin{equation}
\frac{t_d}{\tau} < \frac{t_p^2}{\tau^2}
\end{equation} 
or, simplified,
\begin{equation}
t_p > \sqrt{t_d\tau}.
\end{equation}

Using a standard order of magnitude value for $\tau$ and $t_d$, $\tau = 10^{-8}\,{\rm s}$ ($100\,{\rm MHz}$ sampling) and $t_d = 0.1\,{\rm s}$ we get that $t_p > 10^{-4.5}{\rm s}.$
This means that incoherent dedispersion analysis of pulses shorter than about $10^{-4.5}\,$s usually loses sensitivity relative to coherent dedispersion. 

\section{Existing algorithms for incoherent dedispersion}\label{sec:ExistingMethods}
Algorithmically, there are two leading approaches for inchorerent dedispersion of single-dish data streams.
These are the the tree dedispersion \citep{TreeDedispersion} and brute force dedispersion (e.g., \citealp{GPUsub-bandDedispersion};\citealp{GPUDedispersion}; \citealp{FPGADedispersion}).

The tree dedispersion algorithm, efficiently calculates the integrals of all the straight line paths with slopes between $45^\circ$ and $90^\circ$ through the input time vs. frequency matrix (this is similar to the discrete Radon transform, \citealt{RadonGotsDruckmuller,RadonBrady}).
The computational complexity of this algorithm is $N_{t}N_{f}\log_2{N_{f}}$,
where we use the notation $N_{t} = N_s/N_{f}$ (note that $N_t$ and $N_f$ are the dimensions of $I(t,f)$).
However, since the dispersion curve is not linear, the use of this method
results in a substantial loss of sensitivity.
This can be somewhat mitigated by applying the algorithm to many small sub-bands\footnote{A sub-band of the data is a part of the data that is both limited and continuous in frequency.} of the data, and then combining the results with a dedicated algorithm. This approach is not exact, and it increases the computational complexity of the algorithm. For more details on the sensitivity analysis and computational complexity of this algorithm we refer readers to \citet{GPUDedispersion}.

The brute force dedispersion algorithm simply scans all the trial 
dispersion measures, one at a time, integrating its path on the input map and finding curves with excess power.
This method is exact, but has the high complexity of $N_{\Delta}N_{f}N_{t}$,
where $N_{\Delta}$ is the number of trial dispersion measures scanned (note that $N_{\Delta}={N_{d}}/{N_{f}}$).
In order to expedite the search speed, the algorithm was implemented on graphical processing units (GPUs),
and this method is now capable of analyzing single beam data in real time \citep{GPUDedispersion}.
The maximal sensitivity, along with the possibility of real time analysis using GPUs, makes this method likely to be the most popular algorithm for dedispersion.

Here we present an algorithm that combines both maximal sensitivity and low computational complexity.
A comparison of all these mentioned algorithms is summarized in Table \ref{AlgorithmComparisonTable}.
 
\begin{deluxetable*}{lccc}[H]  
\tablecolumns{4}
\tablecaption{Algorithm comparison \label{AlgorithmComparisonTable}}
\tablehead{   
  \colhead{} &
  \colhead{FDMT} &
  \colhead{Brute force} &
  \colhead{Tree}
}
\startdata
Computational complexity & $\max\{N_tN_{\Delta}\log_2(N_f),2N_fN_t\}$ & $N_fN_tN_{\Delta} $& $N_fN_t\log_2(N_f)$ \\
Information efficient    & Yes & Yes  & No \\
Memory access friendly   & Yes & Yes & Yes \\
Parallelization friendly & Yes & Yes & No 
\enddata
\tablecomments{Comparison of the FDMT algorithm with two other approaches to incoherent dedispersion, brute force (e.g.,\citealp{GPUDedispersion}), and tree dedispersion \citep{TreeDedispersion}. It is clear that the FDMT algorithm dominates in all parameters.}
\end{deluxetable*}

\section{The FDMT Algorithm}\label{sec:FDMTAlgorithm}
The input to the FDMT algorithm is a two dimensional array of intensities, denoted by $I(t,f)$.
The FDMT algorithm calculates the integral over all curves defined by Equation \ref{eq:DtD}.
A dispersion curve can be uniquely defined by the arrival time of the signal at the lowest frequency ($t_0$) and the time delay between the arrival times of the lowest and highest frequencies ($\Delta t$). 

Therefore, the FDMT result can be expressed as
a two dimensional array that contains the integrals
along dispersion curves as a function of
$t_{0}$ and $\Delta{t}$

\begin{equation}\label{parametrization}
A_{f_{\min}}^{f_{\max}}(t_0,\Delta t) = \sum_{f=f_{\min}}^{f_{\max}}{I(t_0 -d\left[\frac{1}{f_{\min}^2} - \frac{1}{f^2}\right],f)},
\end{equation}
where $f_{\rm min}$ and $f_{\rm max}$ are the minimum and maximum frequencies in the base-band. 

To compute the FDMT transform of the input, the algorithm works in $\log_2(N_f)$ iterations. 
The inputs of the $i$'th iteration are the FDMT transforms of a partition of the original band into $N_f/2^{(i-1)}$ sub-bands of size $2^{(i-1)}$ frequencies.  
The outputs of $i$'th iteration are the FDMT transforms of a partition of the original input into $N_f/2^i$ sub-bands of size $2^{i}$ frequencies. Every two successive sub-bands are combined using the addition rule described below. 
After $\log_2(N_f)$ iterations, we have the FDMT over the whole band.
\begin{figure*}
\centering
\includegraphics[width = 180mm]{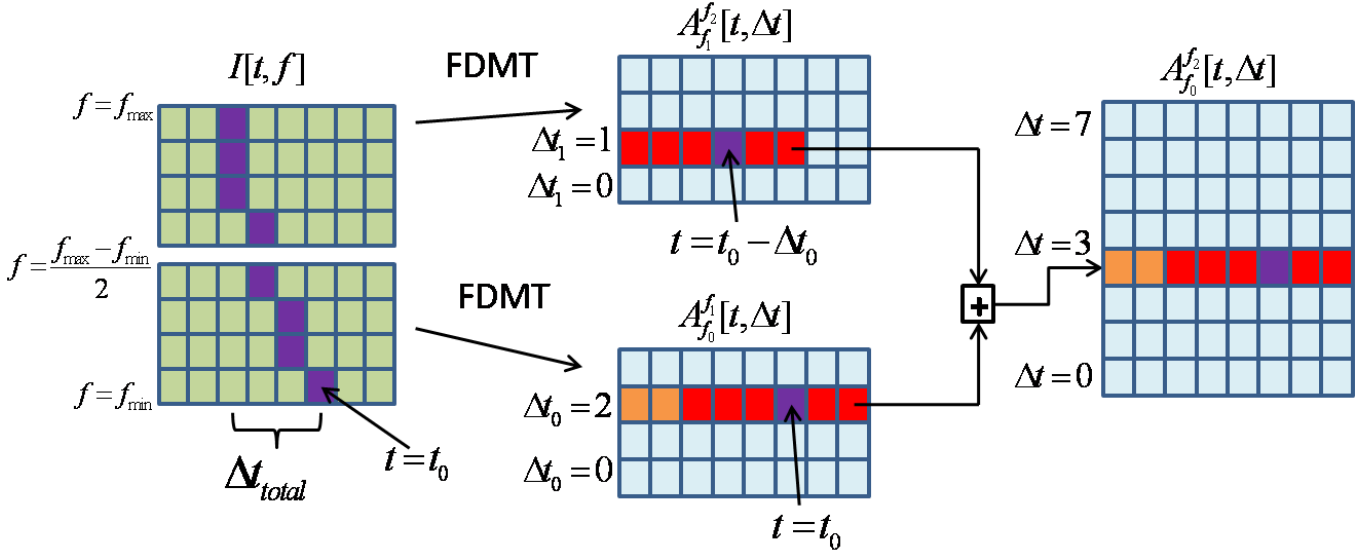}
\caption{Illustration of a single iteration of the FDMT algorithm: On the left side, the input frequency vs. time input table is drawn. An example dispersion curve is highlighted on purple. The input table is divided methodically to two sub-bands.  The right table shows the final dedispersion transform of the input (left), where the integral over the purple dispersion curve (left) is marked by the purple cell on the right. On the middle, the dispersion measure transform of the two sub-bands (which are assumed to be calculated in the previous iteration) is drawn, where each of the two sub-bands contains in each cell the sum of the unique dispersion trail with exit point $t_0$ and total delay $\Delta t$ through the corresponding sub-bands. The cells that contain the partial sums of the two halves of the purple dispersion curve on the left are highlighted in purple. Highlighted in red, is a row in the dispersion tables that contributes to the calculation of the red cells on the right. Notice that we can add the lines highlighted in red as vectors, in order to implement the algorithm in a vectorized form. Highlighted in orange, are the cells that use the alternative addition rule, in the case when the dispersion trail exits the boundaries of the input table.  }\label{fig:AdditionRule}
\end{figure*}

The FDMT combining process of two successive sub-bands into $A_{f_0}^{f_2}(t_0,\Delta t)$, is given by the following addition rule:
\begin{equation}
A_{f_0}^{f_2}(t_0,\Delta t) = A_{f_0}^{f_1}(t_0,t_0 - t_1) + A_{f_1}^{f_2}(t_1 ,t_1-t_2).
\end{equation}
Here, $A_{f_0}^{f_1}$ and $A_{f_1}^{f_2}$ are part of the output of the previous iteration and $t_1$ is the intersection time  of the dispersion curve at the central frequency $f_1 = \frac{f_2-f_0}{2}$.
$t_1$ is uniquely determined by the formula
\begin{equation}
t_1 \equiv t_0 - \Delta t\frac{f_1^{-2} - f_0^{-2}}{f_2^{-2} - f_0^{-2}} \equiv t_0 - C_{f_2,f_0}\Delta t,
\end{equation}
where 
\begin{equation} \label{Eq:Cf2f0}
C_{f_2,f_0} \equiv \frac{f_1^{-2} - f_0^{-2}}{f_2^{-2} - f_0^{-2}}.
\end{equation} 

By definition, $A_{f_0}^{f_1}(t_0,t_0-t_1)$ is the calculated sum over the unique dispersion curve between the coordinates $(t_0,f_0)$ and $(t_1,f_1)$, and $A_{f_1}^{f_2}(t_1,t_1-t_0)$ is the same for $(t_1,f_1)$ and $(t_2,f_2)$.
After an FDMT iteration, the only dispersion curve passing through $(t_0,f_0),(t_2,f_2)$, will be given by $A_{f_0}^{f_2}(t_0,t_0-t_2)$.

For sufficiently early $t_0$, the time $t_1$ will be smaller than zero. In that case we just copy -- i.e, use the alternative addition rule
\begin{equation}
A_{f_0}^{f_2}(t_0,\Delta t) = A_{f_0}^{f_1}(t_0,t_0 - t_1).
\end{equation}
The operation of one iteration of the algorithm is graphically illustrated in Figure \ref{fig:AdditionRule}.

The only thing left to deal with is the data initialization.
This is done prior to the first iteration, generating 
$ A_{f_0}^{f_1}$ for every two consecutive frequencies.
If the maximum dispersion delay between two consecutive frequencies is smaller than the width of a time bin, then we can use the simple initialization:
\begin{equation}
 A_{f}^{f + \delta_f}(t_0,0) = I(f,t),
\end{equation}
where \begin{equation}\delta_f = \frac{f_{\max} - f_{\min}}{N_f}.\end{equation}
Otherwise, the energy of the signal at some frequencies is not located at a single bin. This can be compensated for by two ways:
\begin{enumerate}
\item By computing partial sums over the time axis. i.e 
\begin{equation}
A_{f}^{f + \delta_f}(t,\Delta t) = \sum_{i=0}^{\Delta t}{I(f,t+i)}.
\end{equation}
\item If for a certain $d$, the time delay within each single frequency bin is larger than one time bin, we can simply reduce the time resolution (i.e., bin).
Note that this implies $N_\Delta>>N_f$, which we show in \S \ref{sec:SensitivityIncoherentDedispersion} to be suboptimal in terms of sensitivity.
\end{enumerate}
In the MATLAB and Python codes we provide, we use Option 1.
The maximal time delay within each frequency bin is uniquely determined by $d_{\max}$, the maximal $d$ we want to scan, and is given by
\begin{equation}
\Delta t_{\max}(f_0) =  d_{\max}\frac{f_0^{-2} - (f_0 + \delta_f)^{-2}} {f_{\min}^{-2} - f_{\max}^{-2}}.
\end{equation}
Therefore, this decision can be made prior to the computation.
A pseudo code of FDMT is given in Algorithm \ref{FDMT}.
In addition, we provide implementation for the algorithm in Python and MATLAB.\footnote{The codes are available from https://sites.google.com/site/barakzackayhomepage/home}
Note that so far we did not treat rounding and binning issues, these are discussed in \S \ref{subsec:ImplementataionDetailsFDMT} and are implemented in the codes we provide.
\begin{algorithm}[H]
\caption{The FDMT Algorithm}
\label{FDMT}
Input: $I(f,t)$ input matrix (possibly packed), $t$ axis is continuous in memory.

Output: Packed dispersion measure scores arranged in a two dimensional table  $A_{f_{\min}}^{f_{\max}}(t,\Delta t)$ where $\Delta t$ represents the dispersion measure axis.
\begin{algorithmic}
  \STATE Initiate the table by  $ A_{f}^{f + \frac{f_{\max} - f_{\min}}{N_f}}(t,\Delta t) = \sum_{i=0}^{\Delta t}{I(f,t+i)}$
  
  \FOR{iteration $i=1$ to $i=log_2{N_f}$}
  \FOR{$f_0$ in the range $\left[f_{\min},f_{\max}\right]$ with steps $2^i\delta_f$}
   
  \STATE $f_2 = f_0 + 2^i\delta_f$
  \STATE $f_1 = \frac{f_2 - f_0}{2}$
  \STATE $C_{f_2,f_0} = \frac{f_1^{-2} - f_0^{-2}}{f_2^{-2} - f_0^{-2}}$
  \STATE $\Delta t_{\max}(i,f_0) = N_\Delta \frac{f_0^{-2} - (f_0 + 2^i\delta_f)^{-2}} {f_{\min}^{-2} - f_{\max}^{-2}} $
  \FOR{$\Delta t$ in the range $[0,\Delta t_{\max}(i,f_0)]$}
  \FOR{$t_0$ in range $[C_{f_2,f_0} \Delta t,N_t]$}
  \STATE $t_1 = t_0 - C_{f_2,f_0}\Delta t$
  \STATE
  $
  A_{f_0}^{f_2}(t_0,\Delta t) = A_{f_0}^{f_1}(t_0,t_0-t_1) + A_{f_1}^{f_2}(t_1,t_1-t_2)
  $
  \ENDFOR
  \FOR{$t_0$ in the range $[0,C_{f_2,f_0} \Delta t]$}
    \STATE
    $
    A_{f_0}^{f_2}(t_0,\Delta t) = A_{f_0}^{f_1}(t_0,C_{f_2,f_0} \Delta t)
    $
    \ENDFOR

  \ENDFOR
  \ENDFOR 
  \ENDFOR
  
\end{algorithmic}
\end{algorithm}

\subsection{Computational Complexity}
To calculate the computational complexity, we need to trace the number of operations done throughout the algorithm.
The amount of additions in iteration $i$ is bounded from above by $N_bN_t\Delta t_{\rm max}(i)$ where $N_b$ is the number of sub-bands processed at the current iteration, and $\Delta t_{\max}(i)$ is the maximum time shift within a single sub-band at iteration $i$ for the curve with highest dispersion measure:
\begin{equation}
\Delta t_{\max}(i,f_0) = N_\Delta \frac{f_0^{-2} - (f_0 + 2^i\delta_f)^{-2}} {f_{\min}^{-2} - f_{\max}^{-2}},
\end{equation}

\begin{align}
\Delta t_{\max}(i) = \max_{f_0}\{\Delta t_{\max}(i,f_0)\} = \\ \nonumber = N_\Delta \frac{f_{\min}^{-2} - (f_{\min} + 2^i\delta_f)^{-2}} {f_{\min}^{-2} - f_{\max}^{-2}}.
\end{align}

As a first approximation, one can assume that the dispersion curve is almost linear, meaning that the number of $\Delta t$'s needed in iteration $i$ is roughly twice the number needed in iteration $i+1$.
In the last iteration, $|\Delta t| = N_\Delta$, and therefore, in iteration $i$, 
\begin{equation}
\Delta t_{\max}(i) \approx \frac{N_\Delta2^i}{N_f}.
\end{equation}
The number of bands ($N_b$) in each iteration is $N_b = \frac{N_f}{2^i}$. Therefore, under the approximation of almost linear dispersion (or narrow band), the following approximation is correct:
\begin{equation}
N_b\Delta t_{\max}(i) \approx \max\{N_\Delta, N_b\}
\end{equation}
Summing this for all iterations, and assuming $N_\Delta$ is dominant in all iterations,  and taking into account the number of entries in each added row ($N_t$), we get the complexity 
\begin{equation}
C_{\rm FDMT} = N_tN_\Delta\log_2(N_f).
\end{equation}
If we assume $N_b$ is dominant, we get \begin{equation}C_{\rm FDMT} = N_tN_f +\frac{N_tN_f}{2} + ... = 2N_tN_f.\end{equation}
Therefore, the total complexity of the algorithm is bounded from above by:
\begin{equation}
C_{\rm FDMT} \leq 2N_tN_f + N_tN_\Delta\log_{2}(N_f).
\end{equation}

Casting the complexity analysis of the FDMT algorithm with the more naturally defined $N_s,N_d$, using $N_s = N_fN_t$ and $N_\Delta = N_d/N_f$ we get \begin{align}C_{\rm FDMT} \leq 2{N_{f}}\frac{{N_{s}}}{{N_{f}}} + \frac{{N_{d}}}{{N_{f}}}  \frac{{N_{s}}}{{N_{f}}}  \log_2({N_{f}}) =\\ \nonumber = 2{N_{s}} + \frac{{N_{s}}{N_{d}}}{{N_{f}}^2}\log_2({N_{f}}).\end{align}
Adding to the above the complexity of data preparation by STFT, ${N_{s}}\log_2({N_{f}})$, and the fact that if we chose ${N_{f}}<{N_{p}}$ we can effectively bin (or low pass, see \S \ref{subsec:FDMTFFTPulseProfile}) to size ${N_{p}}$, we get
\begin{equation}C_{\rm FDMT} \leq {N_{s}}\log_2{{N_{f}}} + \frac{2{N_{f}}{N_{s}}}{{N_{p}}} + \frac{{N_{d}}{N_{s}}}{{N_{p}}^2}\log_2({N_{f}}).\end{equation}
Here we can see that the data preparation complexity dominates the operation count of the algorithm whenever incoherent dedispersion is maximally sensitive (i.e, ${N_{d}} < {N_{p}}^2$).

\subsection{Implementation details}\label{subsec:ImplementataionDetailsFDMT}
In this subsection we consider implementation issues, such as rounding and binning, pulse profile convolution and using an arbitrary number of frequencies.
In addition, it is important to implement the tricks of the trade, in order to transform the theoretical complexity reduction to a real speedup. 
\newline {\bf Rounding and binning:}
The exact formulas written above need to take into account discreteness of both frequency and time axes. To keep the formulas readable, we did not include these considerations in the algorithm description and pseudo-code. However, we include them in the implementation we provide, and we advise readers who want to implement the FDMT to pay attention to the discretization process. That is because an incorrect choice might lead to a significant reduction in accuracy.

An example of the most important discretization issue, is that when combining two sub-bands, the point $t_1$, where the dispersion curve travels from one sub-band to the next might not be well defined. This can happen, because the dispersion curve might travel one bin between the end frequency of the first band $f_0+(2^i-1)\delta_f$ and the start frequency of the second band $f_0 + 2^i\delta_f$.
The implemented solution for this problem is to calculate two versions of $C_{f_0,f_2}$, one with the end frequency of the lower sub-band, and the other with the start frequency of the upper sub-band.
Using the different versions of $C_{f_0,f_2}$ in the two different uses of $t_1$, we can account for a time shift between the added bands, approximating the dispersion curve better.

{\bf Machine word utilization:}
One can utilize the machine word width (or the width of the SSE registers) to pack few instances of the dedispersion procedure into one computation (since modern computers operate on machine words of 64 bits, this will result in a speedup factor of 4--8 depending on the number of bits per frequency and the pulse maximum allowed strength).

{\bf Memory access:}
An important issue in run-time reduction, is the continuity of memory access. The FDMT algorithm never performs any re-ordering action along the time axis.
Therefore, it is recommended to store the time axis continuously in order to speed up the memory access operations. 

{\bf Different range of dispersions:}
Sometimes, we have a prior knowledge of the range of dispersion measures needed to scan. In that case, one can still employ the FDMT algorithm after an additional preparation of applying either a frequency dependent shift to the input (according to some $d_{\min}$), or a coherent dedispersion of the signal (using $d_{\min}$).

{\bf Pulse profile:}\label{subsubsec:PulseProfile}
Sometimes we have prior knowledge on the pulse width or profile (might be a different profile for each frequency, like in pulse scattering).
By applying a matched filter approach, one can convolve each frequency time series with the predicted profile for that frequency and employ the FDMT at the end. For wide enough pulse shapes, one might consider binning the time resolution.

We note that convolution of the time axis with a uniform pulse profile (for all frequencies) commutes with the entire FDMT operation.
Therefore, we can test a few pulse profiles per FDMT without repeating the dedispersion process.
 
{\bf Dealing with the case of $N_f\neq 2^k$:}
The algorithm presented above assumes that the number of frequencies is strictly a power of two. This assumption can be abandoned by slightly adjusting the addition rule to allow a merge of non-equal size sub-bands. The only change needed is to switch $f_1$ in Equation \ref{Eq:Cf2f0} from being the middle frequency between $f_0$ and $f_2$ to be the border frequency between the sub-bands.

{\bf Applying FDMT for other functional forms:}
The dispersion equation (Eq. \ref{eq:DtD}) is used only in the preparation of $C_{f_2,f_0}$. One can easily extend the FDMT algorithm to search for other functional forms, for example, \begin{equation}\Delta t = f_1^{\gamma} - f_2^{\gamma}.\end{equation}
The only required modification is to change the power of the frequency in Equation \ref{Eq:Cf2f0} from $-2$ to $\gamma$.
Furthermore, it could be extended to any family of curves that satisfies the condition that there is only one curve passing between any two points in the input data. Using this, one can calculate the required $C_{f_2,f_0}$, by finding the only curve passing through both $(t_0,f_0)$ and $(t_2,f_2)$, and defining $t_1$ to be the intersection time of the curve with the frequency $f_1$.
While the complexity of the algorithm, may change with the functional form, for a sufficiently regular functional form, the complexity will close to $N_dN_t\log{N_f}.$

\section{Eliminating shifts by FFT'ing the time axis}\label{sec:FDMTFFT}
In modern computers and GPU's, memory access is frequently the bottleneck of many algorithms, especially when programming transforms, where the computational complexity is only slightly larger than the data size.
Efficient implementation of transform algorithms is non-trivial and requires architecture dependent changes in order to avoid cache misses \footnote{In modern computers, the fastest memory buffer is the L1 cache. An access to a value that is not stored in the L1 cache causes a memory read from slower storage media such as L2 cache or the RAM memory, and is sometimes called a "cache miss".} (in a general CPU setup) or to avoid communication when using distributed computing.

While it is probably possible to control the algorithms behavior as presented above, it is non-trivial to distribute the data between different processing units while avoiding duplication and communication issues.

In this subsection, we present a variant of the algorithm which is easily parallelized on all architectures and where the memory access pattern is as parallelization friendly as possible. 

The algorithm, as it is described in \S \ref{sec:FDMTAlgorithm}, has only one core operation: adding a complete shifted "time" row. It is the shift operation which makes the data transfer and memory management of the algorithm challenging, and therefore we wish to eliminate shifts from the algorithm.
In order to do that, we can Fourier transform the time axis. This makes the shift operation become a multiplication with a "shift vector" which is the Fourier transform of a shifted delta function.
In this version, all additions are of numbers from the same (Fourier transformed) time coordinate. Therefore, we can assign different parts of the (Fourier transformed) time axis to different processing units, and consequently reduce the need for shared memory or data transport. At the end, we need to Fourier transform back the time axis. We call this algorithm FFT--FDMT and it is summarized in Algorithm \ref{FDMTFFT}. 
Tracking the data in this algorithm, we can see that there are only two "global" steps and that they are both transpose operations of the data. To perform all other steps of the algorithm we need only to access memory that is not larger than one row or one column of the input. Since present L1 cache architectures can contain more than a typical row or column of data, the algorithm can be computing-power limited.
The run-time of this algorithm on any machine is comparable to the run-time of two dimensional convolution, because of the similar number of operations and data access patterns.
We note that in the basic preparation of radio data, one often applies Fourier transforms (for example, when applying filters or screening for radio frequency interferences). 
Therefore, if we have the computational ability to prepare the input table from the raw data, FFT--FDMT is also feasible.   

\begin{algorithm}[H]
\caption{The FFT--FDMT Algorithm}
\label{FDMTFFT}
Input: $I(f,t)$ input matrix (possibly packed), $t$ axis is continuous in memory.

Output: Packed dispersion measure scores arranged in a two dimensional table  $A_{f_{\min}}^{f_{\max}}(t,\Delta t)$ where $\Delta t$ represents the "dispersion measure" axis.
\begin{algorithmic}[1]
  \STATE Initiate the table by  $$ A_{f}^{f + \frac{f_{\max} - f_{\min}}{N_f}}(t,\Delta t) = \sum_{i=0}^{\Delta t}{I(f,t+i)}$$
  \STATE Initiate the "shift vector" $V(\tilde{t_0},\Delta T) = \mathcal{F}(\delta(\Delta T))(\tilde{t_0})$ where $\delta(x)$ is a vector containing one at position $x$ and zeros everywhere else, $\mathcal{F}$ is the FFT operator, and $\tilde{t_0}$ is the index of the Fourier transformed time axis. 
  \STATE Fourier transform the time axis $$ B_{f}^{f + \delta_f}(\tilde{t},\Delta t) = \mathcal{F}(A_{f}^{f + \delta_f}(:,\Delta t))$$
  \STATE Transpose the data. after this action, the frequency and $\Delta t$ axes should be continuous in memory, time axis should be distributed across all computing units.
  \FOR{$\tilde{t_0}$ in the range $[0,N_t]$}
  \FOR{ $i$ in the range $[1,\log_2{N_f}]$}
  \FOR{$f_0$ in the range $\left[f_{\min},f_{\max}\right]$ with steps $2^i\delta_f$}
   
  \STATE $f_2 = f_0 + 2^i\delta_f$
  \STATE $f_1 = \frac{f_2 - f_0}{2}$
  \STATE $C_{f_2,f_0} = \frac{f_1^{-2} - f_0^{-2}}{f_2^{-2} - f_0^{-2}}$
  \STATE $\Delta t_{\max}(i,f_0) = N_\Delta \frac{f_0^{-2} - (f_0 + 2^i\delta_f)^{-2}} {f_{\min}^{-2} - f_{\max}^{-2}} $
  \FOR{$\Delta t$ in the range $[0,\Delta t_{\max}(i,f_0)]$}
  \STATE $\Delta t_1 = C_{f_2,f_0}\Delta t$
  \STATE
  \begin{flalign*}
  B_{f_0}^{f_2}(\tilde{t_0},\Delta t) &= \\ B_{f_0}^{f_1}(\tilde{t_0},\Delta t_1) &+ B_{f_1}^{f_2}(\tilde{t_0},\Delta t - \Delta t_1)V(\tilde{t_0},\Delta t_1)
  \end{flalign*}
  \ENDFOR
  \ENDFOR
  \ENDFOR 
  \ENDFOR
  \STATE Transpose the data back. Now, time is again continous in memory.
  \STATE Perform inverse Fourier transform on the time axis. $$ A_{f_{\min}}^{f_{\max}}(\tilde{t},\Delta t) = \mathcal{F}^{-1}(B_{f_{\min}}^{f_{\max}}(:,\Delta t))$$
\end{algorithmic}
\end{algorithm}

\subsection{Comments on the Implementation of the FFT-FDMT algorithm}

The FFT-FDMT algorithm is designed to increase the amount of computation per cache replacement. To completely optimize the algorithm for this property, we have to consider special implementation details like cache size and processing units communication geometry. Though important to an efficient implementation of the algorithm, these details are considered out of scope for this paper as they are architecture dependent.
We note that all the details discussed in \S \ref{subsec:ImplementataionDetailsFDMT} are valid also for the FFT-FDMT version, except for the changes listed below.

{\bf Machine word utilization:}
The long integer data type is the optimal choice for the regular FDMT algorithm in order to fully utilize the machine word capability. In the Fourier transformed version of the algorithm, we have to use the complex floating point data type.
Using the floating point data type, we have to leave unused the bits of the exponent field, and leave some more bits unused to retain the floating point precision needed to perform the Fourier transform operations.  
Furthermore, some architectures such as GPUs, have a clear optimization preference for the 32 bit floating point data type.
However, it is possible to pack another algorithm instance in the complex field of the input vector. Since the result of the FDMT algorithm is real (as a sum of real numbers), packing another input to the imaginary part of the input is possible. The imaginary part of the result will be the second algorithm instance.

{\bf Pulse profile:} \label{subsec:FDMTFFTPulseProfile}
In addition to the ability to test several pulse profiles per FDMT operation, as explained in \S \ref{subsubsec:PulseProfile}, we can further exploit the use of the Fourier transformed time domain. 
If the pulse width is slightly larger than one bin, reducing the computational load by binning loses information.
Instead, we can effectively apply a low-pass filter on the time axis by either keeping less (time-domain) frequencies or multiplying with a filter. This can be both more sensitive than binning the time axis and more efficient than having a high sampling rate.

{\bf Handling Large dispersion measures:}
If the maximum dispersion broadens the pulse to more than one time bin per frequency bin, the initialization phase of the algorithm inflates the data from size $N_tN_f$ to size $N_\Delta N_f$ (note that the use of $N_\Delta >> N_f$ is losing sensitivity, and therefore this part is not considered a crucial part of the algorithm). The partial sum operation of the initialization phase is equivalent to an application of a low-pass filter on the time axis. This allows natural reduction of computation and memory by saving a differential amount of Fouriered time bins for different $\Delta t$'s. This can be used in the case of large dispersion measures to reduce the algorithm's complexity from $N_\Delta N_t\log_2(N_f)$ to $2N_tN_f\log_2(\frac{N_\Delta}{N_f})$.   

{\bf Zero padding:}
Since convolution is a cyclical operation, all the shifts done in this algorithm are cyclical shifts. Therefore, we have to pad the time axis with $N_\Delta$ zeros prior to the Fourier transform. This operation can increase by a factor of two the complexity of the algorithm if $N_\Delta=N_t$. To avoid this we can choose $N_t >> N_\Delta$. This is usually possible if the size of the input table is not too close to the maximum memory (or cache) capacity of the machine used.  

\section{Run time and Benchmarking}\label{sec:RuntimeAndBenchmarking}

Accurate benchmarking of algorithms should use a mature code, and contain architecture dependent adaptations.
However, it is important to demonstrate that the code we present, running on a single standard CPU, is competitive with the brute force dedispersion implementations on GPU's.
Therefore, we provide a simple benchmark for the provided code.

The benchmark we use is the run-time of performing FDMT on data with the following properties:
$N_f = 2^{10}$, $N_t = 5\times2^{16}$ and $N_\Delta = 2^{10}$.
This volume of input is similar to the one used in the "toy observation" defined in \citep{GPUDedispersion,GPUsub-bandDedispersion}, $N_f = 2^8$, $N_t = 2^{19}$ and $N_\Delta = 500$. Although, we modified the partition between $N_f,N_t$, and increased $N_d$ by a factor of two\footnote{This is a more realistic choice, since using large $N_\Delta>N_f$ usually looses sensitivity (see \S \ref{sec:SensitivityIncoherentDedispersion}), and the number of frequencies is usually larger than $2^{10}$.}.
The run-time we achieve on this data is 3.5 seconds, on a standard Intel Core i-5 4690 processor.
For example, these numbers can represent a real time dedispersion of 8 seconds of input data with 40\,MHz bandwidth and 1024 dispersions.
To get this benchmark, we pack five instances of the algorithm to the 64 bit machine word, allocating 12 bits to each instance. The resulting packed data has dimensions $N_t = 2^{16},N_f=2^{10}$, and serves as input to the FDMT implementation. Using this scheme, we find that our run-time is already shorter than that of the state of the art brute force implementations on GPU's reported in \citep{GPUDedispersion,GPUsub-bandDedispersion}. A comparison between the run-times is shown in Table \ref{RuntimeComparisonTable}.

\begin{deluxetable*}{lccc}  
\tablecolumns{4}
\tablecaption{Runtime comparison \label{RuntimeComparisonTable}}
\tablehead{   
  \colhead{} &
  \colhead{This work} &
  \colhead{\cite{GPUsub-bandDedispersion}} &
  \colhead{\cite{GPUDedispersion}}
}
\startdata

Machine used & Intel Core i5 4690 & Tesla C1060 GPU & Tesla C1060 GPU\\
Programming language & Python (anaconda + accelerate) & C & C\\
Number of instances packed & 5 & 1 & 1 \\
Runtime & 3.5s & 4.8s & 2.1s\\
$N_f,N_t(\rm total),N_d$ & $2^{10},5\times 2^{16},2^{10}$ & $2^8,2^{19},500$ & $2^8,2^{19},500$ \\
$N_fN_tN_d$ & $5\times 2^{36}$ & $2^{36}$ & $2^{36}$ \\
Algorithm used & FDMT (non-FFT version) & Brute force & Brute force \\
Algorithm theoretical complexity & $N_tN_f + N_tN_d\log_2(N_f)$ & $N_fN_tN_d$ & $N_fN_tN_d$  
\enddata
\tablecomments{The FDMT algorithm has a different computational complexity scaling than the brute-force dedispersion it is compared to. Even with standard CPU's and with a high-level programming language, the FDMT implementation we present here is faster than existing GPU implementations of brute force dedispersion.}
\end{deluxetable*}

\section{Bridging the gap between coherent and incoherent dedispersion}\label{sec:CoherentIncoherentHybrid}

Since some interesting transient sources such as pulsar giant pulses are in the regime $1<<{N_{p}}<\sqrt{{N_{d}}}$, it is of importance to find a feasible and sensitive algorithm for their exact dedispersion. 
Coherent dedispersion was, until now, the only sensitive alternative.
The noise power summed when searching for a pulse that is dispersed with a dispersion measure $d$ is \begin{equation} N_{p}+ N_d.\end{equation}
The noise power summed when searching for a non-dispersed pulse is $N_p$, and therefore the largest dispersion tolerable for sensitive pulse detection satisfies  \begin{equation}N_{d} = {N_{p}}.\end{equation}
Therefore, for sensitive detection, the number of dispersion measure trials we need to process is  
\begin{equation} \frac{N_d}{N_{p}}.\end{equation}
The convolution operation performed for coherent dedispersion can be efficiently calculated with Fourier transforms of size ${N_{d}}$, and therefore the complexity of coherent dedispersion is:
\begin{equation}
C_{\rm coherent} = \frac{{N_{d}}}{{N_{p}}}{N_{s}}\log_2({N_{d}}).
\end{equation}
Noting that the computational complexity of coherent dedispersion scales with $N_{d}/N_{p}$ and that of incoherent dedispersion scales with $N_{d}/{N_{p}^2}$, we see that using coherent dedispersion is not computationally efficient for resolved pulses (i.e $N_{p}>1$).

\subsection{Hybrid algorithm for dedispersion}
In order to have both the detection sensitivity of coherent dedispersion and the computational complexity of FDMT, we propose the following solution:
Coherently dedisperse the raw signal with coarse trial dispersion values (with steps $\delta{d}$), and then apply STFT and absolute value squared, followed by FDMT with the maximal dispersion being the next coarse-trial coherent dedispersion.
This process ensures that the FDMT will not lose sensitivity,
relative to coherent dedispersion.

We denote by $N_{\delta{d}}$ the number of bins of length $\tau$ that a delta function pulse will spread upon when dispersed by $\delta{d}$:
\begin{equation}
N_{\delta d} = \frac{4.15\delta d(f^{-2}_{\min} - f^{-2}_{\max}){\rm ms}}{\tau}
\end{equation}
As shown in \S \ref{sec:SensitivityIncoherentDedispersion}, in order to retain sensitivity the maximal dispersion residual to be processed by the following FDMT must be bounded from above by  
\begin{equation}
N_{\delta d} = N_p^2
\end{equation}

Therefore, the number of trial dispersions we need to coherently dedisperse is 
\begin{equation} 
N_{\rm coherent}=\frac{N_d}{N_{p}^2}.
\end{equation}

This process is approaching maximum sensitivity, and its complexity is:
\begin{align}
C_{\rm hybrid} = \frac{{N_{d}}}{N_{p}^2}{N_{s}}(\log_2({N_{d}}) + \log_2({N_{f}})) +\\ \nonumber + \frac{2{N_{d}}{N_{s}}}{{N_{p}}^2} + \frac{{N_{d}}{N_{s}}}{{N_{p}}^2}\log_2({N_{f}}). 
\end{align} 
Simplifying, we get the computational complexity for detection of a pulse of length ${N_{p}}$:
\begin{equation}
C_{\rm hybrid} = \frac{{N_{d}}{N_{s}}}{{N_{p}}^2}(2 + \log_2({N_{d}}) + 2\log_2({N_{p}})). 
\end{equation}
This complexity is near optimal, because the number of uncorrelated scores is $\frac{{N_{d}}{N_{s}}}{{N_{p}}^2}$, which is only a logarithmic factor smaller than the computational complexity. Therefore, there is not much room for further reduction of computational complexity. The algorithm is summarized in Algorithm \ref{alg:AlgorithmCoherent}.

\begin{algorithm}[H]
\caption{Coherent hybrid FDMT dedispersion algorithm}
\label{alg:AlgorithmCoherent}
Input: Antenna voltage series $x(t)$.\newline
Output: Score table for all dispersions $d<{N_{d}}$ with steps ${N_{p}}$ and all exit times $t_0<{N_{s}}$ with steps ${N_{p}}$. 

\begin{algorithmic}[1]

\FOR {dispersion $d_0$ in the range $\left[0,d_{\max}\right]$ in steps of  $\frac{{N_{p}}^2}{{N_{d}}}d_{\max}$}
\STATE Create the signal $y(t)$ by applying the filter $\hat{H}_{d_0}(f) = \exp{\left(\frac{2\pi i d_0}{f+f_0}\right)}$ to $x(t)$.
\STATE Apply STFT with block size ${N_{p}}$ on $y(t)$, to obtain $I(t,f)$.
\STATE Apply FDMT to $I$, with maximal $\Delta t_{\max}={N_{p}}$, and output the partial result $A_{f_{\min}}^{f_{\max}}(d_0 + d,t_0)$ for $d<{N_{p}}^2$ with steps ${N_{p}}$, and  $t_0$ in the range $[0,{N_{s}}]$ with steps ${N_{p}}$.
\ENDFOR
\end{algorithmic}
\end{algorithm}

\subsection{Implications}
Using this algorithm, it is possible to perform blind searches for pulses with duration in the $1\,{\rm \mu s}$ -- $1\,$ms regime (which implies ${N_{p}} = 10^2 - 10^5$ for standard searches).
Together with the low computational complexity of FDMT, this can be efficiently employed in blind searches for FRBs and giant pulses, both reducing the computational load, and increasing sensitivity.

\section{FDMT for radio interferometers} \label{sec:DedispersionInterferometers}
In this section, we analyze the complexity of blind searches of short astrophysical signals with radio interferometers using the FDMT. We first calculate the computational and communication complexity of applying the FDMT algorithm after the imaging operation. Afterwards, we offer a way to reduce the communication complexity by applying the FDMT algorithm after the correlator operation and before the imaging operation.
We show that in principle, using our scheme, it is possible to use modern radio interferometers to detect and locate short astrophysical pulses in real-time without the knowledge of the dispersion measure.

We start by introducing some additional notation.
In the scenario of a blind dispersed pulse search with a radio interferometer, we have signals of several telescopes.
We denote the raw voltage signal from the $j$'th telescope by $x_j$.
We further denote by $N_{a}$ the number of antennas, and by $N_{l}$ the number of distinct locations in the sky, or pixels, in the optimal image resolution of the interferometer.
The desired statistic that we need to calculate for efficient detection is given by:
\begin{equation}
S(t_0,p_x,p_y) = \sum_{t = t_0}^{t = t + {N_{p}}}{\sum_{i=0}^{N}{(x_i\otimes H)(t + u_i(p_x,p_y))}}, 
\end{equation}
where $u_i(p_x,p_y)$ represents the time delay of the signal at antenna $i$, $H$ is the dedispersion filter needed to be convolved with to correct for dispersion, and $\otimes$ represents convolution.
We wish to calculate this score for all combinations of sky positions (which we denote their number by $N_l$), dispersions $\frac{{N_{d}}}{{N_{p}}}$, and start times $\frac{{N_{s}}}{{N_{p}}}$. Therefore, the number of calculated scores is $\frac{N_{l}{N_{s}}{N_{d}}}{{N_{p}}^2}$.

We estimate the number of computations required by using general modern radio interferometer parameters such as:
$\nu = 100\,{\rm MHz}$, $t_d = 0.1\,{\rm s}$, $t_p = 0.1\,{ms}$. Using $N_a = 300$ antennas of diameter $10\,{\rm m}$, spread out to a maximal baseline of $10\,{\rm km}$.
$N_{l}=10^6$, $N_{s} = 10^8$, ${N_{p}}=10^4$, ${N_{d}} = 10^7$ we get $10^{13}$ scores per second, which requires a computing power of $10\,{\rm TFlop/s}$ to process.
The computational requirement of the solution we propose in the next section is only logarithmically larger than this computation rate.
Therefore, it is feasible with current facilities to perform a blind search using modern radio interferometers. 
For example, the computing facility of the Australian Commonwealth Scientific and Industrial Research Organisation\footnote{http://www.atnf.csiro.au/} (known as CSIRO) has computing power equivalent to 260 TFlop/s \footnote{Taken from the Top500 website, http://www.top500.org/list/2014/06/?page=2}.

\subsection{The standard approaches to pulse blind search with interferometers}
There are two existing approaches to blind search interferometry.
The first is to add antennas incoherently and then dedisperse.
This process loses the angular resolution and reduces the sensitivity by a factor of $\sqrt{N}$. However, this is considered to be computationally feasible, and it is sensitive to the interferometer's entire field of view.

The second approach is to "beam-form" and dedisperse, i.e for every searched location $(p_x,p_y)$, shift all the signals from all antennas with the correct shift for position $(p_x,p_y)$, add them up, and perform dedispersion.
To mitigate the computational load of this process, it is custom to use only a small subset of all $N_l$ sky locations at a time, considerably reducing the overall survey speed of the instrument.

Another possibility is to use a combination of both approaches by dividing the interferometers to closely packed stations, beam-forming all stations to a subset of all possible directions, and then incoherently adding the stations.
All methods trade the computational un-feasibility with a significant sensitivity reduction.
A consideration of those approaches can be found in \citep{LOFARPulsar}.
\subsection{The proposed solution}
First, we quickly review the standard imaging process of interferometry, using the approximations of flat sky and short observation. Assuming there is no dispersion, the desired score is 
\begin{align}
S(t_0,p_x,p_y) = \sum_{t = t_0}^{t = t_0 + {N_{p}}}{\left|\sum_{j=0}^{N}{x_j(t + u_j(p_x,p_y))}\right|^2} \\  \nonumber
=\sum_{f=f_{\rm min}}^{f_{\rm max}}{\left|\sum_{j=0}^{N}{\hat{x}_j(t_0,f) \exp(-2\pi ifu_j(p_x,p_y))}\right|^2}\\ \nonumber
=\sum_{f=f_{\rm min}}^{f_{\rm max}}{\sum_{j,k=0}^{N}{\hat{x}_j(t_0,f)\hat{x}_k^*(t_0,f)}}\times\\ \nonumber \exp\left(-2\pi if(u_j(p_x,p_y) - u_k(p_x,p_y))\right),
\end{align}
denoting by $\hat{x}$ the Fourier transform of $x$.
To efficiently calculate this score for every pixel $p_x,p_y$, it is useful to use the relation 
\begin{equation}
u_j(p_x,p_y) - u_k(p_x,p_y) \propto (L_j-L_k)\cdot (p_x,p_y),
\end{equation}
denoting by $L_j$ the two dimensional location of antenna $j$ on the plane (under the approximation of having all antennas on the same plane).
Now, we can calculate the score at all positions $(p_x,p_y)$ at the same time, by a two dimensional fourier transform of the array:
\begin{equation}\label{Eq:InterferometerFXScore}
\hat{S}(t_0,p_u,p_v) = \sum_{j,k,f}{\hat{x}_j(t_0,f)\hat{x}_k(t_0,f)\mathbb{1}((L_j-L_k)f,(p_u,p_v))}
\end{equation}
\begin{equation}
S(t_0,p_x,p_y) = \mathcal{F}^{-1}(\hat{S}(t_0,p_u,p_v)),
\end{equation}
where $\mathbb{1}((a,b),(c,d))$ is equal to one if $(a,b)=(c,d)$ (to the desired approximation), and zero otherwise.

The summation in Equation \ref{Eq:InterferometerFXScore} is a sum of squares. This means that coherent dedispersion operations must be performed before correlating\footnote{The process of calculating $\hat{x}_i(t_0,f)\hat{x}_j(t_0,f)$ is referred to as "correlating" in the literature, and is calculated by a computing infrastructure usually called "the correlator".}, because the imaging process calculates the sum of square absolute values of frequencies.

Incorporating dedispersion into this, we can see that the block size ${N_{f}}$ we used earlier is transformed in this framework to the size of the Fourier transform done by the correlator. As a result, the imaging process cannot be done simultaneously in all frequencies, as different frequency sets should be used for different dispersion measures.
Naively, this means that we need to image separately at each frequency, performing many two dimensional Fourier transform imaging operations, followed by an FDMT for every pixel.
Denoting the complexity of the $i$'th step of the algorithm by $C_i$, the complexity of the coherent dedispersion + STFT of all individual antennas is
\begin{equation}
C_1 = \max\left(1,\frac{{N_{d}}}{{N_{p}}^2}\right)N_aN_s(\log_2({N_{d}}) + \log_2(N_f)).
\end{equation}
The complexity correlating all pairs of antennas is
\begin{equation} C_2 = \max\left(1,\frac{{N_{d}}}{{N_{p}}^2}\right)\frac{N_a(N_a-1)}{2}{N_{s}}. \end{equation}
The complexity of the imaging process is  
\begin{equation} C_3 =  N_l\log_{2}(N_l)\frac{{N_{s}}{N_{d}}}{{N_{p}}^2}.  \end{equation}
The complexity of the FDMT algorithm (without the STFT part which was already done in this context) is 
\begin{equation} C_4 = N_l\frac{{N_{s}}{N_{d}}}{{N_{p}}^2}\log_{2}({N_{f}}).
\end{equation}
So, the total complexity of this process is
\begin{equation}
C = C_1 + C_2 + C_3 + C_4.
\end{equation}

While the complexity of this process is indeed "optimal", in the sense that it is only a logarithmic factor larger than the number of independant results, implementing this will result in a reduced computational efficiency. This is due to data transport between the imaging stage and the dedispersion stage.
Between these stages, $\frac{N_l{N_{s}}{N_{d}}}{{N_{p}}^2}$ complex numbers are being transported.

This could be mitigated by the fact that dedispersion can be done before the imaging operation, if the condition 
\begin{equation}\label{eqn:NoChromatic}
(f_{\rm max} - f_{\rm min}) (u_i(p_u,p_v)-u_j(p_u,p_v)) < 1
\end{equation}
holds\footnote{sometimes, if the Condition \ref{eqn:NoChromatic} doesn't hold, the resulting image is said to suffer from a "chromatic aberration".}.
If the band is wide, this condition will not hold for pairs of far away antennas.
In this case, it is necessary to split the frequencies into sub-bands that are narrow enough to maintain Condition \ref{eqn:NoChromatic}.
Since the FDMT's input and output dimensions have the same size, the communication complexity of the proposed solution is $\frac{N_a(N_a-1){N_{s}}{N_{d}}}{2{N_{p}}^2}$, which should (if $N_a^2 << N_l$) make the algorithm's run-time be computation limited, and thus feasible.

Another interesting point, is that if we are in the regime of ${N_{\Delta}}<{N_{f}}$, then the FDMT is volume shrinking, and performing it only after the imaging process will result in excessive computation in the imaging stage. 
This scenario is sometimes plausible, for example, when looking for pulsars in a globular cluster, where we sometimes have a relatively good guess of the dispersion measure, or if we are using the choice of ${N_{f}}>\sqrt{{N_{d}}}$ (with some sacrifice of sensitivity, if ${N_{f}}>{N_{p}}$), 

This process is summarized in Algorithm \ref{InterferometersAlgorithm}.
\begin{algorithm}[H]
\caption{Finding short pulses with interferometers}
\label{InterferometersAlgorithm}
Input: Antenna voltage series
Output: $S(d,t_0,p_x,p_y)$ for every time, dispersion, and sky location.
Standard choice of ${N_{f}}$ is ${N_{p}}$.
\begin{algorithmic}[1]

\FOR {dispersion $d_0$ in the range $[0,{N_{d}}-\frac{{N_{d}}}{{N_{f}}^2}]$ in steps of $\frac{{N_{d}}}{{N_{f}}^2}$}
\FOR {antenna index $i$.}
\STATE Create the signal $x_i$ by convolving the $i'th$ antenna signal with the dedispersion filter with index $d_0$.
\STATE Apply STFT with block size of ${N_{f}}$ on the signal $x_i$, to obtain $\hat{x}_i$
\ENDFOR
\FOR {every pair of antennas $i,j$ calculate $\hat{x}_i^d\hat{x}_j^d$}
\FOR {each populated point on the $\hat{S}(t_0,f,p_u,p_v)$ matrix} 
\STATE Generate the time vs. frequency map of all the frequencies\footnotemark that enter into the same cell $(p_u,p_v)$.
\STATE Apply FDMT with maximal ${N_{d}}={N_{f}}^2$
\ENDFOR
\STATE Data "transpose operation", each processing unit now holds all the $p_u,p_v$ $\hat{S}(t_0,d,p_u,p_v)$ cells, for the same $d,t_0$. 
\FOR {each dispersion $d<{N_{f}}^2$ and each time $t_0$}
\STATE Perform two dimensional inverse Fourier transform to calculate $S(t_0,d,p_x,p_y) = \mathcal{F}^{-1}(\hat{S}(t_0,d,p_u,p_v))$
\STATE If for some point, the power is statistically significant. 
\ENDFOR
\ENDFOR

\ENDFOR 
\end{algorithmic}
\end{algorithm}
\footnotetext[10]{Also, it is possible for certain geometric configurations that several antennas will contribute to the same cell in the U,V plane}

%
%
%

\section{Conclusion}\label{sec:Conclusion}
We present the FDMT algorithm, that performs exact incoherent dedispersion transform with the complexity of $N_fN_t\log_2(N_f)$. We show that regular implementation tricks of the trade can be combined with the FDMT algorithm to achieve significant computation speedup. We also present a variant of the FDMT algorithm that is slightly more computationally intensive, but concentrates all memory access operations to two global transpose operations, and might present further speedup on massively parallel architectures such as GPUs.
We show that the FDMT algorithm dominates all other known algorithms for incoherent dedispersion and has comparable complexity to the signal processing operations required to generate its input data. Therefore, we conclude that incoherent dedispersion can now be considered a non-issue for future surveys.
We provide implementations of the FDMT algorithm in high level programming languages, with a faster runtime than the state of the art implementations of brute-force dedispersion on GPUs. 

We further present an algorithm that bridges the gap between coherent and incoherent dedispersion, and show that the computational complexity of this algorithm is orders of magnitude lower than that of coherent dedispersion for pulses of resolvable duration. Using this algorithm, it will be possible to perform blind searches for FRBs and giant pulse emitting pulsars with the sensitivity of coherent dedispersion searches.

Last, we compute the operation count for a blind search of short astrophysical searches with modern radio interferometers and arrive to the conclusion that it is computationally feasible using existing facilities.    

\section*{acknowledgments}
BZ would like to express his deep thank to Gregg Halinnan for introducing him the problem of dedispersion.
The authors would like to express their thanks to Avishay Gal-Yam, Shrinivas Kulkarni, Matthew Bailes, David Kaplan, Ora Zackay and Gil Cohen for their useful comments and advice regrading the paper.
E.O.O. is incumbent of
the Arye Dissentshik career development chair and
is grateful to support by
grants from the 
Willner Family Leadership Institute
Ilan Gluzman (Secaucus NJ),
Israeli Ministry of Science,
Israel Science Foundation,
Minerva and
the I-CORE Program of the Planning
and Budgeting Committee and The Israel Science Foundation.


\begin{thebibliography}{}
\expandafter\ifx\csname natexlab\endcsname\relax\def\natexlab#1{#1}\fi


\bibitem[Barsdell et al.(2012)]{GPUDedispersion} Barsdell, B.~R., Bailes, M., Barnes, D.~G., \& Fluke, C.~J.\ 2012, \mnras, 422, 379  

\bibitem[Brady (1998)]{RadonBrady}
 Brady, Martin L. "A fast discrete approximation algorithm for the Radon transform." SIAM Journal on Computing 27, no. 1 (1998): 107-119.

\bibitem[Clarke et al.(2013)]{FPGADedispersion} Clarke, N., Macquart, J.-P., \& Trott, C.\ 2013, \apjs, 205, 4 

\bibitem[Clarke et al.(2014)]{DedispersionInterferometers} Clarke, N., D'Addario, L., Navarro, R., \& Trinh, J.\ 2014, Journal of Astronomical Instrumentation , 3, 50004 


\bibitem[Gotz \& Druckmuller(1996)]{RadonGotsDruckmuller}	Gotz, W. A., and H. J. Druckmuller. "A fast digital Radon transform. An efficient means for evaluating the Hough transform." Pattern Recognition 29, no. 4 (1996): 711-718.


\bibitem[Manchester et al.(1996)]{sub-bandDedispersion} Manchester, R.~N., 
Lyne, A.~G., D'Amico, N., et al.\ 1996, \mnras, 279, 1235


\bibitem[Magro et al.(2011)]{GPUsub-bandDedispersion} Magro, A., Karastergiou, 
A., Salvini, S., et al.\ 2011, \mnras, 417, 2642


\bibitem[McLaughlin 
\& Cordes(2003)]{DetectingPulsarsByGiantPulses} McLaughlin, M.~A., \& Cordes, J.~M.\ 2003, \apj, 596, 982 

\bibitem[Lorimer et al.(2007)]{LorimerBurst} Lorimer, D.~R., Bailes, 
M., McLaughlin, M.~A., Narkevic, D.~J., 
\& Crawford, F.\ 2007, Science, 318, 777

\bibitem[Lorimer 
\& Kramer(2012)]{HandbookOfPulsarAstronomy} Lorimer, D.~R., \& Kramer, M.\ 2012, Handbook of Pulsar Astronomy, by D.~R.~Lorimer , M.~Kramer, Cambridge, UK: Cambridge University Press, 2012,

\bibitem[Petroff et al.(2014)]{PetroffFRBsNullDetection} Petroff, E., van 
Straten, W., Johnston, S., et al.\ 2014, \apjl, 789, L26  

\bibitem[Taylor(1974)]{TreeDedispersion} Taylor, J.~H.\ 1974, \aaps, 15, 367 

\bibitem[Thompson et al.(2011)]{DedispersionInterferometersCaseStudy} Thompson, D.~R., Wagstaff, K.~L., Brisken, W.~F., et al.\ 2011, \apj, 735, 98 

\bibitem[Thornton et al.(2013)]{FRBS} Thornton, D., 
Stappers, B., Bailes, M., et al.\ 2013, Science, 341, 53

\bibitem[van Leeuwen \& Stappers(2010)]{LOFARPulsar}
 van Leeuwen, J., \& Stappers, B.~W.\ 2010, \aap, 509, A7


\end{thebibliography}
\end{document}